\documentclass[12pt]{article}
\usepackage{graphicx}
\newcommand\pubnumber{CERN-TH-2018-008}
\newcommand\pubdate{\today}

\def\institute{Theoretical Physics Department, CERN, Geneva, Switzerland, and\\ INFN, Sezione di Milano Bicocca, Milano, ITALY}

\def\Title#1{\begin{center} {\Large #1 } \end{center}}
\def\Author#1{\begin{center}{ \sc #1} \end{center}}
\def\Address#1{\begin{center}{ \it #1} \end{center}}

\newcommand\pubblock{\rightline{\begin{tabular}{l} \pubnumber\\
         \pubdate  \end{tabular}}}
\newenvironment{Abstract}{\begin{quotation}  }{\end{quotation}}
\newenvironment{Presented}{\begin{quotation} \begin{center} 
             PRESENTED AT\end{center}\bigskip 
      \begin{center}\begin{large}}{\end{large}\end{center} \end{quotation}}

\def\beq{\begin{equation}}
\def\eeq#1{\label{#1}\end{equation}}
\def\eeqn{\end{equation}}

\def\beqa{\begin{eqnarray}}
\def\eeqa#1{\label{#1}\end{eqnarray}}
\def\eeqan{\end{eqnarray}}

\let\bar=\overbar

\def\Dslash{\not{\hbox{\kern-4pt $D$}}}
\def\dslash{\not{\hbox{\kern-2pt $\del$}}}

\def\mt{m_t}

\def\msb{{\bar{\ssstyle M \kern -1pt S}}}

\newcommand\hvq{$hvq$}
\newcommand\ttbnlodec{$t\bar{t}dec$}
\newcommand\bbfourl{$b\bar{b}4\ell$}
\newcommand\MSbar{\ensuremath{\rm \overline{MS}}}
\newcommand\mwbj{\ensuremath{m_{Wb_j}}}
\newcommand\mwbjmax{\ensuremath{m_{Wb_j}^{\max}}}
\newcommand\Pythia{{\tt Pythia}}
\newcommand\Herwig{{\tt Herwig}}
\newcommand\PythiaEight{{\tt Pythia8}}
\newcommand\HerwigSeven{{\tt Herwig7}}
\newcommand\HerwigSevenPone{{\tt Herwig7.1}}

\newcommand\PythiaEightPtwo{{\tt Pythia8.2}}
\newcommand\HerwigSevenPlot{{\tt He7.1}}
\newcommand\PythiaEightPlot{{\tt Py8.2}}

\begin{document}
\begin{titlepage}
\pubblock

\vfill
\Title{The Top Quark Mass at the LHC}
\vfill
\Author{Paolo Nason}
\Address{\institute}
\vfill
\begin{Abstract}
  I briefly discuss some theoretical aspects of top mass measurements at the LHC. In particular, I illustrate a recent theoretical study performed using next-to-leading order (NLO) calculations interfaced to shower generators (NLO+PS) of increasing accuracy, interfaced to both Pythia8 and Herwig7 Monte Carlo generators.
\end{Abstract}
\vfill
\begin{Presented}
$10^{th}$ International Workshop on Top Quark Physics\\
Braga, Portugal,  September 17--22, 2017
\end{Presented}
\vfill
\end{titlepage}
\def\thefootnote{\fnsymbol{footnote}}
\setcounter{footnote}{0}

\section{Introduction}
An important goal of the LHC top-physics program is the
measurement of its mass. Since the Higgs mass is
known with high precision, improvements of both the $W$ and top mass
measurements may lead to a refinement of the Electro-Weak
precision tests~\cite{Patrignani:2016xqp-EWreview, Baak:2014ora}.
The current precision of the
$W$ mass measurement, of about 15~MeV, would match
a precision on the top mass of about 2.4~GeV.

There is some tension at present between the value
of the top mass obtained indirectly through electro-weak
fits~\cite{Patrignani:2016xqp-EWreview}, $176.7\pm 2.1$~GeV,
and the direct determinations, with the value of
$173.34\pm 0.76$~GeV from the latest combination~\cite{ATLAS:2014wva},
and with later measurements yielding values smaller
by about 1~GeV~\cite{Aaboud:2016igd, Khachatryan:2015hba,%
CMS-PAS-TOP-17-007, ATLAS-CONF-2017-071}.\footnote{%
For recent reviews of top-mass measurements by the
ATLAS and CMS collaborations see Refs.~\cite{Pearson:2017jck} and
\cite{Castro:2017yxe} from these proceedings.}

The value of the top mass is also relevant for
the issue of vacuum stability in the Standard
Model~\cite{Degrassi:2012ry,Buttazzo:2013uya, Andreassen:2017rzq}.
Direct measurements are now well below the instability region,
while the central value extracted from electro-weak fits
is near its edge. The only conclusion that one can
draw from these results is that no indication of new physics
scales below the Plank scale arises from the vacuum
metastability requirement. On the other hand, the very small
value of the Higgs quartic coupling near the Planck scale
is an intriguing coincidence, even if at the moment we do not
know how to interpret it.

The relatively small errors on top mass measurements quoted by the
experimental collaborations has been challenged in some theoretical
works, that claimed that the mass extracted in direct
measurements is not related to a well defined field-theoretical mass
parameter. This claim has appeared in different forms, and with
different meanings depending upon the authors. In
ref.~\cite{Hoang:2008xm} it is argued in essence that the difference
between the pole mass and the Monte Carlo mass parameter is due to effects of
non-perturbative origin, and to effects of order $\alpha_s\Gamma_t$.
Other publications claim that since the Shower Monte
Carlos used to extract the top mass have only leading order accuracy,
they cannot be possibly sensitive to a well defined field theoretical
mass like the \MSbar{} or the pole mass, since they start to differ at
next-to-leading order accuracy\cite{Alioli:2013mxa}.
Yet in other works it is argued that the use of jets should be avoided
in top mass measurements, since those are affected by hadronization
errors~\cite{Kawabataa:2014osa}.
Several theoretical proposals of alternative methods to measure the top mass
have appeared in the literature, sometimes motivated by the objections
listed above~\cite{Alioli:2013mxa,Kawabataa:2014osa,Hoang:2017kmk,
  Agashe:2016bok,Frixione:2014ala,Kawabata:2016aya}.
Furthermore, experimental results are often separated into ``direct
measurements'' and ``pole mass measurements'', where the latter are
obtained by comparing experimental measurements with calculations performed
at least at the next-to-leading order level, and no qualification is
given to what kind of mass parameter is measured in direct measurements.

It has also been argued that the pole mass is not a viable mass
parameter for top mass measurements, because of the mass renormalon
problem~\cite{Hoang:2008xm}. Recent studies, however, have shown that
the renormalon ambiguity is safely below the current experimental
errors, being equal to 110~MeV according to
ref.~\cite{Beneke:2016cbu}, and to 250~MeV according to
ref.~\cite{Hoang:2017btd} (for a critical discussion of the larger
uncertainty obtained there, see ref.~\cite{Nason:2017cxd}).

In ref.~\cite{Nason:2017cxd} I have argued that direct measurement
should be considered pole mass measurements. In short, it is easy to
argue that this is the case as far as perturbation theory is
concerned, and non-perturbative effects can be estimated in the usual
way using Monte Carlo hadronization models, with special attention to
their aspects that are particularly worrisome in top mass measurements
(as for the case of colour
reconnection~\cite{Wicke:2008iz,Sjostrand:2013cya}).  Furthermore,
there are recent implementation of NLO calculations interfaced to
parton shower generators~\cite{Campbell:2014kua,Jezo:2016ujg} that are
particularly relevant for studying whether subtle perturbative effects
can have important consequences in top mass measurements, and are typically
implemented in the (complex) pole mass scheme.

In ref.~\cite{Ravasio:2018lzi} we have performed a study using recent generators
for top production, aimed at estimating theoretical errors in top mass
measurements. We have considered three generators of increasing accuracy:
the \hvq{} generator~\cite{Frixione:2007nw}, that implements NLO corrections
  only in production, and is widely used by the experimental
collaborations in top-mass analyses; the \ttbnlodec{}~\cite{Campbell:2014kua}
generator, that also implements NLO corrections in top decay and exact
spin correlations in the narrow width approximation, and the \bbfourl{}~\cite{Jezo:2016ujg}
generator, that also implements finite width and non-resonant contributions,
including interference effects of radiation in production and decay.

We have focused our study on a simplified observable, the mass of a ``particle level
top'' defined as the system made up of the hardest lepton, the hardest neutrino, and
the jet containing the hardest $B$ meson, all with the appropriate flavour
to match a top or an anti-top. The peak of this mass distribution, that we call
\mwbjmax{}, is of course strongly correlated with the input top mass, that corresponds
to the pole mass scheme, since this is the scheme adopted in the NLO calculations
of the three generators. Our aim was then to examine the dependence of \mwbjmax{}
on the generator being used (and also on parameters settings, like the
factorization and renormalization scale in each generator) for the same input top mass.
Since a differences in \mwbjmax{} would result in a difference in the value
of the extracted top mass of nearly the same magnitude and opposite sign
when examining the same data set, we are in a position to determine intrinsic
errors due to parameter settings, and errors due to the use of the less
accurate generators.

The result of the comparison of the three generators interfaced
to \PythiaEightPtwo{}~\cite{Sjostrand:2014zea} is reported in table~\ref{tab:mwbj_showerOnly}.
\begin{table*}[htb]
\centering
%
\begin{tabular}{l|c|c|c|c|}
  \cline{2-5}
 &  \multicolumn{2}{ |c|}{PS only}
 &  \multicolumn{2}{ |c|}{ \phantom{\Big|} full}\\
 \cline{2-5}
 & \phantom{\Big|} No smearing & smearing
 & \phantom{\Big|} No smearing & smearing \\
 \cline{1-5}
 \multicolumn{1}{ |c|  }{ \phantom{\Big|}  \bbfourl{}}
 & $172.522$~GeV
 & $171.403$~GeV
 & $172.793$~GeV
 & $172.717$~GeV
 \\ \cline{1-5}
 \multicolumn{1}{ |c|  }{ \phantom{\Big|}\ttbnlodec{} ${}-$ \bbfourl{}}
  &  $         -18 \pm            2 $~MeV
  &  $+         191 \pm            2 $~MeV
  &  $+          21 \pm            6 $~MeV
  &  $+         140 \pm            2 $~MeV
 \\ \cline{1-5}
 \multicolumn{1}{ |c|  }{ \phantom{\Big|}\hvq{} ${}-$ \bbfourl{}}
  &  $         -24 \pm            2 $~MeV
  &  $         -89 \pm            2 $~MeV
  &  $+          10 \pm            6 $~MeV
  &  $        -147 \pm            2 $~MeV
 \\ \cline{1-5}
\end{tabular}
\caption{Differences in the \mwbjmax{} for $\mt$=172.5~GeV for
  \ttbnlodec{} and \hvq{} with respect to \bbfourl{}, showered with
  \PythiaEightPtwo{}, at the NLO+PS level and at the full hadron level.
  Results obtained after smearing the \mwbj{} distribution with a Gaussian
  function with a 15~GeV width are also shown in order to mimic effects
  due to experimental uncertainties.}
\label{tab:mwbj_showerOnly}
\end{table*}
Besides reporting the ``bare'' \mwbjmax{} value,
we also report the \mwbjmax{} value obtained after the application of a
Gaussian smearing to the \mwbj{} distribution,
with a Gaussian width equal to 15~GeV (which is the typical
experimental resolution of the reconstructed top mass)
in order to mimic detector resolution effects.
From the table we see that the shift in the peak position is very small
for the bare distribution, while it is of the order of 100~MeV in the
smeared case.

The very good agreement among the three generators may seem
strange at first sight, since the \hvq{} generator does not implement
NLO correction to radiation in top decay, and this radiation may
influence the peak position, since it controls how much energy is capture in
the jet cone. It is however understandable if we remember that \Pythia{}
implements Matrix Element Corrections (MEC) in top decay, and in our case these
are equivalent to NLO accuracy. If MEC are switched off we see a variation
of $-61$~MeV in the bare \mwbjmax{} for the \hvq{} generator, while the variation
becomes close to -1~GeV for the smeared distribution. This is understood as
being due to the fact that the peak position is dominated by events
where most radiation in decay is captured by the jet, while when smearing is
performed, events that fall on the left side of \mwbjmax{}, associated
to large angle radiation in decay, also contribute.

In ref.~\cite{Ravasio:2018lzi} several other sources of errors are considered, but
none of them is disturbing, leading to the conclusion that the improvement brought
by the new generators, and in particular the inclusion of off-shell, non-resonant
contribution and the interference of radiation in production and decay, do not
displace the peak of the reconstructed mass by more than about 150~MeV.

A very disturbing result is instead found if \HerwigSeven{}~\cite{Bahr:2008pv, Bellm:2015jjp} is used,
as can be seen in table~\ref{tab:mass_extraction-shower}.
\begin{table*}[htb]
\centering
\begin{tabular}{l|c|c|c|c|}
 \cline{2-5}
 &  \multicolumn{2}{ |c|}{ \phantom{\Big|} No smearing}
 &  \multicolumn{2}{ |c|}{15~GeV smearing} \\
 \cline{2-5}
 & \phantom{\Big|} {\HerwigSevenPlot} & {\PythiaEightPlot} ${}-$ {\HerwigSevenPlot} 
 & \phantom{\Big|} {\HerwigSevenPlot} & {\PythiaEightPlot} ${}-$ {\HerwigSevenPlot}\\
 \cline{1-5}
 \multicolumn{1}{ |c|  }{ \phantom{\Big|}\bbfourl{}}
 & $172.727$~GeV 
  &  $+          66 \pm            7 $~MeV
 & $171.626$~GeV 
  &  $+        1091 \pm            2 $~MeV
 \\ \cline{1-5}
 \multicolumn{1}{ |c|  }{ \phantom{\Big|}\ttbnlodec{}}
 & $172.775$~GeV 
  &  $+          39 \pm            5 $~MeV
 & $171.678$~GeV 
  &  $+        1179 \pm            2 $~MeV
 \\ \cline{1-5}
 \multicolumn{1}{ |c|  }{ \phantom{\Big|}\hvq{}}
 & $173.038$~GeV 
  &  $        -235 \pm            5 $~MeV
 & $172.319$~GeV 
  &  $+         251 \pm            2 $~MeV
 \\ \cline{1-5}
\end{tabular}
\caption{$\mwbj{}$ peak position for $\mt$=172.5~GeV obtained with the three
  different generators, showered with \HerwigSevenPone{}~({\HerwigSevenPlot}). The
  differences with \PythiaEightPtwo{}~({\tt Py8.2}) are also shown.}
\label{tab:mass_extraction-shower}
\end{table*}
In this case the \hvq{} generator differs substantially from \bbfourl{} and
\ttbnlodec{} even for the bare \mwbj{} distribution, where it exceeds \bbfourl{}
by more than 300~MeV, and even more for the smeared one, where the excess
raises to almost 700~MeV. Furthermore, the difference between \PythiaEight{}
and \HerwigSeven{} for the smeared distribution when using the \bbfourl{} and
\ttbnlodec{} generators is larger than 1~GeV.
In the \hvq{} case the difference is of the order of 250~MeV and of opposite
sign in the bare and smeared case. This signals that
the relatively small 250~MeV difference in the smeared case
is the accidental consequence of cancellation
effects due to the very different description of the reconstructed mass peak
in the two Monte Carlos.

In ref.~\cite{Ravasio:2018lzi} we also examined other observables, namely the
peak of the $b$-jet energy~\cite{Agashe:2016bok} and the set of
leptonic observables considered in ref.~\cite{Frixione:2014ala}.
Also in these cases we found large differences among the \Pythia{} and \Herwig{}
results. In the case of the leptonic observables, this finding contrasts with the
naive expectation that leptonic observables should be insensitive to shower
and hadronization effects.

It is unlikely that the 1~GeV difference found between
\Pythia{} and \Herwig{} may
translate directly  into a corresponding top mass uncertainty in
realistic analysis.\footnote{In \cite{Ravasio:2018lzi} it is also found that the dependence
of \mwbjmax{} as a function of the jet radius is different in the two Monte Carlos,
and thus it is unlikely that both of them may represent the data fairly.}
It is, however, an important issue to be understood, since \Pythia{} and \Herwig{}
differ considerably in the shower model (that is a dipole shower in the
former, and an angular ordered parton shower in the latter). Assuming that no
specific problems are found either in the two Monte Carlos or in their
NLO+PS interfaces, and that both models may be tuned to fit fairly
observables that are relevant for top mass measurements, we would be forced
to consider remaining differences among the two Monte Carlos as sources of
theoretical errors to be accounted for.


\providecommand{\href}[2]{#2}\begingroup\raggedright\endgroup

\end{document}